\documentstyle[preprint,aps]{revtex}

\begin{document}
\title{Probabilistic exact cloning and probabilistic no-signalling}
\author{Arun Kumar Pati}
\address{Quantum Optics and Information Group,}
\address{SEECS, Dean Street, University of Wales, Bangor LL 57 IUT, UK}

\date{\today}

\maketitle

\def\ra{\rangle}
\def\la{\langle}
\def\ver{\arrowvert}

\begin{abstract}
We show that non-local resources cannot be used for {\em probabilistic
signalling} even if one can produce {\em exact clones}
with the help of a probabilistic quantum cloning  machine (PQCM).
We show that  PQCM cannot help to distinguish two
statistical mixtures at a remote location. Thus
quantum theory rules out the possibility of sending superluminal signals
not only deterministically but also {\em probabilistically}.
We give a bound on the success probability of producing multiple clones
in an entangled system.

\end{abstract}

\bigskip


All the useful information that can be transmitted has a universal speed limit, 
namely, the speed of light. 
In quantum mechanics the situation seems to be changed when 
Einstein, Podolsky and Rosen put forward their famous thought experiment on 
two systems that had interacted in the past but are no longer in direct 
contact \cite{epr}. These kind of systems are described by entangled states
which show non-local correlations that cannot be explained by any local
hidden variable theory \cite{jb}. In \cite{epr} they proved that the
measurement outcome of one system can instantaneously affect the result of 
 the other. This suggested that one can exploit the non-local nature of the 
entangled states to send superluminal signals. However, it was shown that since 
the operators at space-like separated distance commute, the averages of the 
observable at distant site remain the same and do not depend on the
operations carried out by the other (distant) party \cite{pe,grw}. 

Subsequently, Herbert \cite{nh} put forward the idea of amplifying quantum 
states on one part of the system and performing a single measurement on many 
copies to get information about what the other party has done. It was then 
demonstrated by Wooters-Zurek \cite{wz} and Dieks \cite{dd} that such a 
multiplying device cannot exist due to linearity of quantum theory. This is 
now known as the ``no-cloning theorem''. Exact cloning of an unknown quantum
state allows
one to copy one part of an entangled system many times, which would give the 
possibility to find out what observable the other party has measured. This 
provides us with the means to communicate instantaneously. Since deterministic
and exact cloning is ruled out by linear process, the superluminal
signalling using non-local nature of entangled states was doomed.
Further, it was found that it is the unitarity of quantum evolutions \cite{hy}
which prevents us from cloning
two non-orthogonal states exactly and this has been used against measuring
a single quantum state \cite{ahy}. In recent years various cloning machines
which produce inaccurate copies \cite{bh,brub} by unitary processes are conceived, but they 
cannot be used for superluminal signalling. In fact, Gisin \cite{ng} has
shown that  the fidelity criterion based on ``no-signalling'' and
inaccurate copying are consistent with each other. The no signalling constraint
has been used to generate optimal asymmetric clones \cite{gkr}.

A recent proposal by Duan and Guo \cite{dg} shows that two non-orthogonal 
states can be cloned {\em exactly} using unitary and measurement
operations with a postselection.
Furthermore, they \cite{dg1} have shown that a set of states chosen secretly can be
{\em exactly cloned} by a {\em probabilistic} cloning machine if and only if 
the states are linearly independent.
Hardy and Song \cite{hs} have used  no-signalling condition for
the probabilistic cloning machine
to find a limit on the number of state that can be cloned in 
a given Hilbert space of a quantum system.
The exact quantum cloning have been
studied from a state discrimination view by Chefles and Barnett \cite{chef}.
They have also studied the interpolation between the inaccurate and exact cloning
and proposed a network to implement the cloning operation \cite{chef1}.
Recently, we  \cite{akp} have
shown that the unitarity of quantum theory allows
us to have a linear superposition of multiple {\em exact} clones along with
a failure term iff the non-orthogonal states chosen secretly are linearly
independent. This ``novel cloning'' machine is quite general one and all the
probabilistic and deterministic cloning machines can be regarded as a special
case of the ``novel cloning'' machine. We have also proposed a protocol
which produces exact clones of an arbitrary qubit (universal cloning) with
the help of classical communication and the use of entanglement \cite{arun}.

It is known that if the quantum states can be cloned exactly then superluminal
signals can be sent definitely. 
The question we address in this letter is
whether exact probabilistic cloning allows us to send {\em
superluminal signals
probabilistically}. If a message can be transmitted instantaneously
with certain {\em non-zero
probability} less than unit, we call it probabilistic superluminal signalling.
We show that given a probabilistic cloning machine which produces exact clones
one cannot send
superluminal signals probabilistically. Whereas deterministic superluminal signalling 
is impossible, there was some doubt that quantum mechanics might admit
probabilistic superluminal
signalling with the help of a probabilistic cloning machine.
We give a bound on the 
success rate of producing $M$ exact clones of one part of system
using PQCM in a composite system. This result shows
that again the quantum theory is in agreement with the principles
of special relativity.

Suppose we have a singlet state consisting of two particles shared by
Alice and Bob. The state is given by

\begin{eqnarray}
&& |\psi^- \ra =  \frac{1}{\sqrt{2}} (\ver 0~1 \ra - \ver 1~0 \ra ) \nonumber \\
&& =  \frac{1}{\sqrt{2}} (\ver \psi \ra \ver \psi_{\bot} \ra- \ver \psi_{\bot} \ra \ver \psi \ra ),
\end{eqnarray}
where $\ver \psi \ra = \alpha \ver 0 \ra + \beta \ver 1 \ra $ and
 $ \ver \psi_{\bot} \ra = \beta^* \ver 0 \ra -  \alpha \ver 1 \ra $
 are mutually orthogonal spin states (or polarisations in case of photon).
We call this basis $\{ \ver \psi \ra, \ver \psi_{\bot} \ra \}$ as the
qubit basis.
Alice is in possession of particle 1 and Bob is in possession of particle 2. 
Alice can chose to measure the spin along the $x$- or the $z$-axis.
It is known that given
a shared EPR state between Alice and Bob, then the measurement outcomes of Bob
are invariant under arbitrary local unitary transformation done by Alice. This
is the basis for no-superluminal communication.
The measurement outcomes of Bob is 

\begin{eqnarray}
 p(a) = tr_B ( \rho_B P_a),
\end{eqnarray}
where the set of operator $\{ P_a \}$ are projectors or could be some
generalised measurements such as POVMs satisfying
 $\sum P_a^{\dagger} P_a = I$. The
density matrix $\rho_B = tr \rho_{AB} = tr [ (U_A \otimes I_B) \rho_{AB}
(U_A \otimes I_B)^{\dagger} ]$
is invariant under a unitary operation by Alice. Hence Bob cannot
distinguish two statistical mixtures resulting at his location due to
the unitary operation done at a remote place. However, if
Bob can produce exact clones (of any arbitrary input state)
of his particle then he can distinguish two
statistical mixtures. Is the same true with a probabilistic cloning machine?

For a single quantum system the probabilistic quantum cloning machine (PQCM)
takes an input state  $|\psi_i\ra$ from a set ${\cal S}=
\{|\psi_i\ra \} (i= 1,2,..,K)$. This state is going to be cloned. Let $|A \ra = \ver 0 
\ra^{\otimes  M}$ an ancilla (a collection of $M$ blank states) and $|P_0\ra,
\ver P_1 \ra$
be the `check' states which after a measurement tells us whether cloning has been 
successful or not. Following \cite{dg1} it can be proved that there is 
a unitary transformation $U$ such that the following evolution holds

\begin{eqnarray}
&& |\psi_i\ra|A\ra|P_0\ra ~\rightarrow U(|\psi_i\ra|A\ra|P_0\ra ) = \nonumber \\
&& ~ \sqrt p_i |\psi_i \ra^{\otimes M} |P_0\ra +
 \sqrt{1-p_i} |\Phi_i \ra |P_1\ra\,
\end{eqnarray}
if and only if the states $\{ \ver \psi_i \ra \}$ are linearly independent.
Here, $p_i$ is the probability of successful cloning $M$ states
and $|\Phi_i \ra$ is the composite state of the input and blank states and
these states need not be orthonormal.
The unitary evolution together with a projection measurement yields $1
 \rightarrow M$ {\em exact clones} with a probability of success $p_i$. 
To investigate the question of the possibility of
the probabilistic superluminal signalling we carry
out the action of PQCM on one part of a composite system (say) on Bob's
particle 2. Bob attaches ancillas $C$ and $D$ and the evolution of the
combined state is given by

\begin{eqnarray}
&& |\psi^- \ra_{AB} |A \ra_C |P_0 \ra_D ~\rightarrow U_{BCD} (|\psi^- \ra_{AB}|A\ra_C|P_0 \ra_D )= \nonumber \\
&& ~\frac{1}{\sqrt 2} ( \sqrt p_{\bot i} \ver \psi_i \ra_A |{\psi}_{\bot i} \ra^{\otimes M}_{BC} -
  \sqrt{ p_i} \ver {\psi}_{\bot i} \ra_A \ver \psi_i \ra^{\otimes M}_{BC} )|P_0 \ra_D \nonumber\\
&& + \frac{1}{\sqrt 2}  ( \sqrt{1-p_{\bot i} } \ver \psi_i \ra_A |\Phi_i \ra_{BC}
- \sqrt{1-p_i } \ver {\psi}_{\bot i} \ra_A \ver {\tilde \Phi_i} \ra_{BC} )|P_1
  \ra_D ,
\end{eqnarray}
where the set $\{ \ver \psi_i \ra \}$ and $\{ \ver {\psi}_{\bot i} \ra \}, (i=1,2)$
are linearly independent. The index $i$ refers to two possible choices of basis
onto which Alice might do a measurement. For example, $\{ \ver \psi_i \ra \} = \{ \ver 0 \ra,
\cos \theta \ver 0 \ra + \sin \theta \ver 1 \ra \}$ and $\{ \ver {\psi}_{\bot i} \ra \} =
\{ \ver 1 \ra, \sin \theta \ver 0 \ra -  \cos \theta \ver 1 \ra \}$.
Bob performs a measurement on the probing device $P$ of the cloning machine.
He keeps the states if the outcome is $\ver P_0 \ra$ and discards the result 
if the outcome is $\ver P_1 \ra$. From Eq.(4) it is clear that 
after postselection of measurement result the antisymmetric Bell state becomes

\begin{eqnarray}
&& \frac{1}{\sqrt 2}( |\psi_i \ra \ver {\psi}_{\bot i} \ra - |{\psi}_{\bot i} \ra \ver \psi_i \ra ) ~\rightarrow ~\nonumber \\
&& \frac{1}{\sqrt 2}( \sqrt{ p_{\bot i} } |\psi_i \ra \ver {\psi}_{\bot i} \ra^{\otimes M} -
 \sqrt{ p_i } |\psi_{\bot i} \ra \ver \psi_i \ra^{\otimes M}) .
\end{eqnarray}
If Bob can follow the above procedure, then the after Alice finds her particle
in the basis $\{ \ver 0 \ra, \ver 1 \ra \}$ the reduced density matrix of the
particle 2 (with ancillas) will be 

\begin{equation}
\rho_{BC} = tr_{DA} (\rho_{ABCD} ) = \frac{1}{2}( p_0 \ver 0 \ra \la 0 \ver^{\otimes M} +
 p_1 \ver 1 \ra \la 1 \ver^{\otimes M} ).
\end{equation}
On the other hand  if Alice finds her particle in the basis
$\ver \psi_2 \ra = \cos \theta \ver 0 \ra + \sin \theta \ver 1 \ra,
 \ver \psi_{\bot 2} \ra = \sin \theta \ver 0 \ra -
 \cos \theta \ver 1 \ra \}$,  then the reduced density matrix of the joint
 system ($BC$) would be

\begin{equation}
\rho_{BC} = tr_{DA}( \rho_{ABCD} ) = \frac{1}{2}( p_2 \ver \psi_2 \ra \la \psi_2 \ver^{\otimes M} +
p_{\bot 2} \ver \psi_{\bot 2} \ra \la \psi_{\bot 2} \ver^{\otimes M} ).
\end{equation}
Since the two statistical mixtures in (6) and (7) are different
this would have allowed Bob to distinguish two preparation stages by Alice, thus
allowing for
superluminal signalling probabilistically. However, this is not possible. Since
there are four states involved in the cloning process and Bob's particle belong
to a two-dimensional Hilbert space only two of them can be linearly independent.
It can be seen from the form of equation (5) that $\{ \ver \psi_2 \ra \}$ and
$\{ \ver \psi_{\bot 2} \ra \}$ must be linearly dependent on the states
$\{ \ver \psi_1\ra \}$ and $\{ \ver \psi_{\bot 1} \ra \}$ . This means Bob's
cloning machine should be able to clone the states from the set 
$\{ \ver 0 \ra, \ver 1 \ra,
\cos \theta \ver 0 \ra + \sin \theta \ver 1 \ra ,
\sin \theta \ver 0 \ra -  \cos \theta \ver 1 \ra \}$. But this
set is linearly dependent, hence Bob cannot clone all these four states exactly
with a non-zero probability.
Thus there is no way for Bob to make an inference based on the measurement results
of his $M$ clones and to know
what Alice has done. Thus there {\em cannot} be any superluminal signalling even
with a {\em non-zero probability less than unit}. 

    The no-signalling constraint leads to the observation by
Hardy and Song \cite{hs} that one cannot clone
more than certain number of linear independent states via a PQCM. Here,
we derive a
bound on the success rate of producing $M$ clones for $N$ linearly
independent states for a composite system. Let us consider an arbitrary
composite system whose state is described by

\begin{equation}
|\psi \ra_{AB} = \sum_{ij=1}^{N_A N_B} a_{ij}\ver x_i \ra \ver y_j \ra,
\end{equation}
where $\{ \ver x_i \ra \} \in {\cal H}_A = C^{N_A}$ and $\{ \ver y_j \} \in
{\cal H}_B = C^{N_B}$. Using the Schimdt decomposition we can write the bipartite
state as

\begin{equation}
|\psi \ra_{AB} = \sum_{k=1}^{N} c_k \ver \alpha_k \ra \ver \beta_k \ra,
\end{equation}
where we have put $N_A = N$, where $N$ is the dimension of the smallest Hilbert
space.  Though Hilbert space ${\cal H}_B$ has $N_B$ number of linearly
independent states in principle, only $N$ of them can be cloned exactly \cite{hs}.
We can write the bipartite state as \cite{ac}

\begin{equation}
|\psi \ra_{AB} = \frac{1}{\sqrt N} \sum_{k=1}^{N} \ver u_k \ra \ver v_k \ra,
\end{equation}
where $\ver u_i \ra$ are orthogonal basis for ${\cal H}_A$ and $\ver v_i \ra$
are non-orthogonal and {\em linearly independent} basis states for ${\cal H}_B$.
Suppose the subsystem $B$ is passed through a PQCM, then we have

\begin{eqnarray}
&& \ver \Psi \ra_{ABCD} = U_{BCD}( |\psi \ra_{AB} \ver A \ra \ver P_0 \ra ) \nonumber \\
&& = \frac{1}{\sqrt N} \sum_{k=1}^{N}
\biggl[ \sqrt{p_k} \ver u_k \ra \ver v_k \ra^{\otimes M} \ver P_0 \ra +
\sqrt{1 - p_k} \ver u_k \ra \ver \Phi_k \ra \ver P_1 \ra  \biggr]
\end{eqnarray}
After a postselction of measurement result the machine will yield the state

\begin{eqnarray}
\ver \tilde{\Psi} \ra_{ABC} =  \frac{1}{\sqrt N} \sum_{k=1}^{N}
\sqrt{p_k} \ver u_k \ra \ver v_k \ra^{\otimes M} .
\end{eqnarray}

We would like to know what is the success rate of producing $M$ clones of two
distinct states at Bob's site chosen from $N$ linearly independent states.
Let us define a basis in the Hilbert space
${\cal H}_{BCD} $ as  $\ver X_i \ra$, given by (using eq.(11) )

\begin{equation}
\ver X_i \ra = \la u_i \ver \Psi \ra_{ABCD}  = \frac{1}{N} 
\biggl[ \sqrt{p_i} \ra \ver v_i \ra^{\otimes M} \ver P_0 \ra +
\sqrt{1 - p_i} \ver \Phi_i \ra \ver P_1 \ra  \biggr].
\end{equation}
Now taking the inner product of the two distinct basis we can get 

\begin{equation}
N \ver \la X_i \ver X_j \ra \ver  \le 
\sqrt{p_i p_j} \ver \la v_i \ver v_i \ra \ver^{M}  +
\sqrt{(1 - p_i)(1 - p_j) } \ver \la \Phi_i \ver \Phi_j \ra \ver.
\end{equation}
Simplifying the above inequality we have

\begin{equation}
\frac{1}{2}(p_i + p_j) \le {(1 -  N  \ver\la X_i| X_j \ra \ver) \over
(1 -  \ver \la v_i | v_j \ra \ver^M ) }
\end{equation}
The success probability in the case of composite system depends on the
overlap of the actual states being cloned and also on the overlap of two
non-orthogonal states belonging to a larger Hilbert space.  In the limit of
infinite number of cloning (when $M \rightarrow \infty $), this bound approaches some kind of state
discrimination bound for a composite system, given by

\begin{equation}
P = \frac{1}{2}(p_i + p_j) \le (1 -  N  \ver\la X_i| X_j \ra \ver) .
\end{equation}
The maximum total success probability depends on the number of linearly
independent states that can be cloned.

In conclusion, we have shown that even though a probabilistic quantum cloning
machine can produce {\em exact} clones with certain non-zero probability
still it is impossible to send superluminal signals with certain non-zero
probability. Thus
in the quantum theory the no-signalling condition is more stringent than
that of the classical
theory. In classical relativity we do not have a deterministic (with
unit probability) superluminal
signalling. But the quantum theory says that we cannot even have a
{\em probabilistic} (less
than unit)  {\em superluminal signalling}. However, Kent \cite{kent} has shown
that in a different quantum theory (in the context of time neutral
cosmology) there is a superluminal signalling with non-zero probability.\\

I thank P. Kok, S. L. Braunstein and L. M. Duan for useful discussions.
I thank L. Hardy for  useful correspondence. The financial support from ESPRC
is gratefully acknowledged.



\end{document}